# My discussions of quantum foundations with John Stewart Bell

**Marian Kupczynski**

Département de l'Informatique, UQO, Case postale 1250, succursale Hull, Gatineau. QC, Canada J8X 3X7

**\* Correspondence:**
marian.kupczynski@uqo.ca
ORCID: MK, 0000-0002-0247-7289

**Abstract**: In 1976, I met John Bell several times in CERN and we talked about a possible violation of optical theorem, purity tests, EPR paradox, Bell's inequalities and their violation. In this review, I resume our discussions, and explain how they were related to my earlier research. I also reproduce handwritten notes, which I gave to Bell during our first meeting and a handwritten letter he sent to me in 1982. We have never met again, but I have continued to discuss BI-CHSH inequalities and their violation in several papers. The research stimulated by Bell's papers and experiments performed to check his inequalities led to several important applications of quantum entanglement in quantum information and quantum technologies. Unfortunately, it led also to extraordinary metaphysical claims and speculations about retro-causality and quantum nonlocality, which in our opinion John Bell would not endorse today. BI-CHSH inequalities are violated in physics and in cognitive science, but it neither proved the completeness of quantum mechanics nor its nonlocality. Quantum computing advantage is not due to some magical instantaneous influences between distant physical systems. Therefore one has to be *cautious in drawing-far-reaching philosophical conclusions from Bell's inequalities*.

## 1. Introduction

   Several times in 1976, I met John Bell in CERN and we discussed different topics, which were connected to my previous research. This is why, in this introduction, I explain how my research evolved from a domain of Lie groups and high energy particle physics to Bell's inequalities (BI) and their violation.
   I studied in parallel physics and mathematics at Warsaw University. After defending in 1967 my master thesis on contractions of Lie groups and their representations I started to work in the Institute of Theoretical Physics at Warsaw University. It was a decade of the relativistic S matrix, which together with various discrete symmetries and the additive quark model was used as a main tool to study high energy scattering of elementary particles and their strong interactions [1, 2]. S matrix is a linear unitary operator which transforms initial states of scattering process (in-states) represented by vectors $|i>$ in a Fock space into final states (out-states) represented by vectors $|f>$. A probability $P_{if}$ for obtaining a particular final state |f > from an initial state |i > is: $P_{if} = |\langle f|S|i \rangle|^2$. By analogy to wave scattering one defines a scattering operator T: $S = I + iT$ and scattering amplitudes $f_{ab} = <b|T|a>$ for obtaining a final state $|b>$ after interaction of particles in the initial state $|a>$.

In my doctoral thesis, completed in 1971, I studied polarization predictions for scattering amplitudes following from the additive quark model [3, 4, 5]. This model, used to study scattering phenomena led to several good predictions but it was only an abstract algorithm allowing making calculations, which was postulated and could not be derived. This is why I lost interest in this model.

In 1972, I received UNESCO post- doctoral scholarship in the International Center for Theoretical Physics in Trieste (ICTP) and I decided to use my stay in ICTP to analyze epistemological foundations of quantum mechanics (QM) and of quantum field theory (QFT). I could not accept that, quantum probabilities for statistical scatter of experimental outcomes are considered to provide a complete description of individual physical systems. I thought that Einstein was right and it should be possible to prove it. I reproduce below few paragraphs, after minor editing, from a preprint [6] in which I resumed some of my conclusions.

"QM recognized that purity of a quantum ensemble is an important notion. QM concentrated on a preparation stage of an experiment. A system was said to be in a pure quantum state, if it passed by a maximal filter or if a complete set of commuting observables was measured on the system. It was not clear how could one know that a filter was maximal or how could one construct it.

In *axiomatic quantum mechanics* (AQM), initiated by a paper of Birkhoff and von Neumann [7], it was claimed that to each vector in a Hilbert space corresponded a realizable physical state of a physical system and that Hilbert space description was general enough to describe all imaginable future phenomena. This claim was refuted by Bogdan Mielnik in two excellent papers [8, 9], in which he showed that one could imagine infinitely many non-Hilbertian 'quantum worlds'. Inspired by Mielnik's papers I decided to analyze various general experimental set-ups, which could be used to investigate phenomena characterizing ensembles of 'particle-like' beams.

This analysis led us to several conclusions [10]:
1) Properties of beams depend on properties of devices and vice-versa. A beam b is characterized by a statistical distribution of outcomes obtained by passing by all devices $d_i$. A device d is defined by statistical distributions of outcomes it produces for all available beams $b_i$. All observables are contextual and observed physical phenomena depend on the richness of beams and devices.
2) In different runs of experiments we observe beams $b_k$, each characterized by its empirical probability distribution. Only if an ensemble ß of all these beams is a pure statistical ensemble of pure beams, we can associate estimated probability distributions with beams b∈ß and with individual particles members of these beams.
3) A pure ensemble ß of pure beams b is characterized by empirical probability distributions s(r) which remain approximately unchanged:
    (i)     for the new ensembles $ß_i$ obtained from the ensemble ß by the application of the i-th intensity reduction procedure on each beam b∈ß
    (ii)    for all rich sub-ensembles of ß chosen in a random way."

In AQM efforts were concentrated on a search for a set of axioms, reflecting general properties of propositions, which could be said or asked about physical systems. However, most of these axiomatic schemes were inspired by experiments with color and polarization filters on an optical bench. Mielnik demonstrated that Hilbert space description might be too poor [9]. I pointed out that such description might be too rich or inadequate [10]:

"A careful analysis leads us to new definitions of filters, pure ensembles and to the important conclusion that in any considered case Hilbert space description turns out to be possible. However, sometimes data do not allow extraction of transition probabilities in a unique way, so it is more reasonable to abandon Hilbert space description. Another feature which appears in our analysis is the fact that only some vectors and some scalar products in Hilbert space may have a physical meaning. Therefore, in some way, Hilbert space language is too rich.

Too rich language makes possible to explain data, using more or less phenomenological models, without really broadening understanding of them. Successes of such models deepen our belief in a fundamental and unchangeable character of the language used and build a psychological barrier, making a discovery of a new, and more economical and less ambiguous language much more difficult.

All these considerations encouraged us to raise an important question whether Hilbert space language is too rich to explain observed physical phenomena, for example in high-energy elementary-particle physics. A natural question arises: how could we find whether this is the case? Although it is evident now that we cannot assign to all vectors in a Hilbert space physically realizable states of elementary particles and that not all scalar products can be practically measured, yet it does not mean that Hilbert space language is necessarily too rich. Similarly, in classical mechanics not every solution of an arbitrary Newton equation has a practical meaning, and this does not mean that the language of classical physics is inappropriate.

To prove that Hilbert space language is too rich to deal with scattering phenomena of elementary particles, we have to show, for example, that the unitarity of S matrix is violated. We would have to find two initial realizable states $|i_1>$ and $|i_2>$, which in our formalism must be represented by orthogonal vectors, and show that final states $|Si_1>$ and $|Si_2>$ cannot be represented by orthogonal vectors."

In QM and in QFT the unitarity of S matrix is a very important notion, because it is implied by the conservation of probability. The unitarity of S matrix and the assumption: $S = I + iT$ allow proving the Optical Theorem (OT), which relates unmeasurable imaginary part of the elastic scattering amplitude in the forward direction to the total cross-section. OT believed to be equivalent to probability conservation became the irreplaceable tool in high energy particle physics.

In 1973, I realized that one may easily reconcile the probability conservation with the absence of an elastic channel. Strong interactions occur only at very short distances. One can imagine that scattering events split incoherently into two groups. In the first group particles are too far to interact strongly. In the second group all pairs of particles interact strongly and it may happen that there is no elastic channel or at least no elastic scattering in the forward direction. If we follow this 'particle-like' intuition instead of the decomposition $S = I + iT$ we obtain $S = I \oplus S_1$, where $S_1$ is a unitary scattering operator describing strong interactions. The probability is conserved, but OT cannot be derived [11-13]. Since OT was used as a constraint in data analysis, only by chance I discovered published data, obtained without using this constraint. These data dramatically violated OT [14]. Unfortunately one cannot prove directly the violation of OT, because all the extrapolations to the forward direction are in fact unreliable [15].

Fortunately, there is another difference between our and the standard description of scattering phenomena. In the standard approach ensembles of initial states are usually assumed to be pure; in our approach we have a mixed ensemble of initial particle pairs differing, for example, by impact parameters.  There is an important difference between pure and mixed ensembles: any sub-ensembles of a pure ensemble have the same properties but sub-ensembles of a mixed ensemble may differ. This is why I proposed and investigated various *purity tests* [16-18], hoping that experimentalists will start using them.  We cannot control a distribution of impact parameters, but in our model a change in geometry of an experiment (for example: intersecting rings ISR experiment in CERN) might lead to observable effects, which were not supposed to occur according to the standard approach.

Apparently Bell and Eberhard had also some doubts about the validity of OT [19]. Therefore, a possible violation of OT and purity tests was the main subject of my discussions with John Bell.  Bell agreed with me that, a slight shift of beams in ISR might have measurable consequences for measured total cross-sections.  However, he doubted that I might convince experimentalists to check my hypothesis. Another topic of our discussions, were of course EPR paradox and BI. To put our discussion of BI in a precise context, I prepared handwritten notes which I gave to him during our first meeting. When we met next time, Bell gave me back my manuscript and we discussed it.

In this manuscript, I explained that, if hidden variables $\lambda$ describing each pair of 'particles' were couples of bi-valued, strictly correlated, spin functions $S_1$ and $S_2$ on a sphere and outcomes for each pair were values of $S_1(a)$ and $S_2(b)$, then one could not use an integration over a set of these functions as it was done in his original proof of inequalities [20, 21]. Then, one may only try to prove BI using estimates of expectations obtained by averaging sums of products $S_1(a)S_2(b)$ over all pairs in long runs of corresponding incompatible experiments. If expectation value $E(a, b)$ is replaced by its estimate a proof of BI may never be rigorous and at least  error bars have to be included. A rigorous proof could only be given, if pairs of spin functions describing entangled particles were the same in all experimental runs performed using different settings, what is highly improbable due to the richness of an uncountable set of spin functions on a sphere.  Bell agreed with me, but I did not think about publishing this manuscript. Nobody thought in 1976, that the experiments testing validity of BI would be rewarded by a Nobel Prize in 2022.

In 1978, Poland signed an agreement with Morocco for a technical and scientific cooperation and I was chosen to teach in Morocco for 2 years at Ecole Hassania d'Ingenieurs à Casablanca. I was planning to return to Warsaw University after my 2 years in Morocco, but because of the political situation in Poland I stayed in Morocco much longer. After two years in Casablanca, I moved to Rabat and started to teach at Ecole Normale Superieure Takaddoum and at University Mohammed V. Finally I emigrated to Canada in 1986.

I was surprised, that John Bell found me in Morocco and sent me a copy of his letter to Itamar Pitowski, in which he mentioned the discussions we had in 1976. He enquired; whether I published the manuscript I discussed with him. Apparently BI and their violation [22-24] became a hot topic. Pitowski constructed spin functions on a sphere and claimed that a local hidden variable model based on these functions was able to reproduce predictions of QM and violate BI [25, 26]. I did not follow in Morocco these developments and I did not know Pitowski's papers. Bell's letter revived my interest in the foundations of quantum mechanics, in Bell inequalities and their violation.

In section 2, I reproduce handwritten notes I gave to John Bell in 1976.
In section 3, I reproduce the letter I received from him in 1982.
In section 4 I resume my reply to John Bell and subsequent publications.
Section 4 contains few conclusions.

## 2. My manuscript: EPR paradox and Bell Inequalities

A statement that a wave function gives a complete knowledge of an individual micro-system leads to a paradox, if highly correlated two-particle states are considered. For example, if

$$\Psi(x_1, x_2) = \sum_{n=1}^{\infty} \psi_n(x_1) u_n(x_2) = \sum_{m=1}^{\infty} \varphi_m(x_1) v_m(x_2) \qquad (1)$$

where $u_n(x_1)$ and $v_m(x_2)$ are *eigenfunctions* of two non-commuting operators A and B respectively, then after a measurement of A on the second particle a state of the first particle is $\psi_k(x_1)$ but after a measurement of B on the second particle a state of the first particle $\varphi_s(x_1)$. Thus a wave function describing the first particle depends on what we decide to measure on the second distant particle.

We find no paradox, if we accept a statistical interpretation according to which $\Psi(x_1, x_2)$ describes only an ensemble of two particles states prepared in a particular way. The reduced quantum states $\psi_k(x_1)$ and $\varphi_l(x_1)$ describe different sub-ensembles of first particles such that the measurements of A on their distant companions "produced" outcomes $a_k$ and $b_s$ respectively.

If we accept the statistical interpretation, then a pure quantum ensemble is in fact a mixed statistical ensemble, because it contains particles differing by some properties. Therefore, one arrives in a very natural way to the conclusion that in fact quantum mechanics does not provide a complete description of individual physical systems and one should be able to find it out.

One way to prove it, is to perform purity tests. Another is to demonstrate, that in any reasonable hidden variable theory one must necessarily, find some predictions differing from the predictions of quantum mechanics. The proofs of it were given by Bell, Wigner, Clauser and Shimony.

Let us now analyze entangled particles prepared in a spin singlet state:

$$\Psi == \frac{1}{\sqrt{2}} (|+>_\theta|->_\theta - |->_\theta|+>_\theta) \qquad (2)$$

where $0\leq\theta\leq\pi$, $|+>_\theta = \cos\theta|+> + \sin\theta|->$ and $|->_\theta = -\sin\theta|+> + \cos\theta|->$ are "spin up" and "spin down" states respectively, when measured in the direction $(\cos\theta, \sin\theta)$ on the x-y plane.

According to the statistical interpretation of QM, if we measure spin projections in the direction $(\cos\theta_1, \sin\theta_1)$ on particles 1 we obtain the outcome "+" or "-" with a probability ½. If we concentrate on the particles 1 for which the measurement outcomes are "+" then the sub-ensemble of their companions is described by a reduced quantum state $|->_{\theta_1} = -\sin\theta_1|+> + \cos\theta_1|->$. Therefore, one can predict with certainty that spin projections in the direction $(\cos\theta_1, \sin\theta_1)$ measured on these companions will be "-".

Since the same prediction is believed to be correct for any chosen direction $(\cos\theta_1, \sin\theta_1)$, Dirac's interpretation of quantum measurement may not be maintained. According to Dirac a quantum system is in some intermediate state and after interacting with a measuring apparatus, it jumps to some definite state yielding a definite outcome with a probability determined by QM. One may not obtain strictly correlated outcomes of distant measurements, if quantum systems "decide", in irreducibly random way, what to do, when measurements are done.

Therefore, if one wants to build a theory of hidden variables explaining this ideal experiment and reproducing quantum correlations one has to assume, that measurement outcomes are <u>strictly</u> determined by some correlated hidden variables describing entangled 'particles' created by a source.

Having this in mind we reexamine Bell's and Wigner's proofs, that it is impossible to attribute predetermined measurement outcomes of incompatibles observables to individual particles and to reproduce quantum predictions for spin polarization experiments.

Both Wigner and Bell assume that there exists some continuous probability density function on a space V of hidden variables and that there is the one to one correspondence of each hidden state of particles with a point or with a subset in this space V. Both Bell and Wigner consider reasonable to write the integrals:

$$\int_V \rho(\lambda) d\mu(\lambda) = 1,  \quad (3)$$

where $\rho(\lambda)$ <u>does not depend on the measurement settings</u> and is a continuous function beyond a set of measure 0.

We proceed, from the beginning, in a different way. We represent a continuous set of polarization filters by angles φ between their optical axes and the y axis on the (x-y) plane. Each "particle" ( a photon or an electron) is described by a discontinuous function $\Psi(\varphi) = \pm 1$, for $\varphi \in [0,\pi]$.

A set of such functions is uncountable infinite and it is not known how to represent a statistical ensemble of such states by using the integral (3), thus we concentrate on finite samples obtained in polarization experiments.

Let us consider an experiment in which a source is sending entangled photon-pairs to distant measuring stations. Before starting polarization measurements we calibrate our source and check that beam intensity is stable and approximately equal to I. Then, we can study a behavior of subsequent groups of photon-pairs in time periods ΔT. We will have on average N±ΔN pairs in each group, where N=IΔT.

Let us assume that we have two distant measuring stations using 4 pairs of polarization filters (A,B), (A,C), (D,B), D,C) and two detectors on each side. The clicks on the detectors are interpreted as registration of the photons with "spin up" or "spin down" and they are coded as ±1.

In each period of time ΔT, photon pairs may be described by specific pairs of functions $(\Psi_1(\varphi),\pm\Psi_1(\varphi)),\ldots \ldots (\Psi_N(\varphi),\pm\Psi_N(\varphi))$ drawn from an uncountable set of possible functions $\Psi(\varphi)$; where '±' correspond to strict correlation or to strict anti-correlation respectively. Using these functions, we may estimate, for each pair of polarizers, pairwise expectations $E(AB) \approx E_{AB} = \pm \frac{1}{N} \sum_{n=1}^{N} \Psi_n(\varphi_A) \Psi_n(\varphi_B)$. We skip '±' in all the equations below because the obtained inequalities and their violation does not depend on the sign of correlations.

Mimicking Bell-CHSH proof and using $|\Psi_n(\varphi)|=1$ we obtain:

$$|E_{AB} - E_{AC}| = \frac{1}{N}\left|\sum_{n=1}^{N}\Psi_n(\varphi_A)\Psi_n(\varphi_B) - \sum_{n=1}^{N}\Psi_n(\varphi_A)\Psi_n(\varphi_C)\right| =$$

$$\frac{1}{N}\left|\sum_{n=1}^{N}\Psi_n(\varphi_A)\Psi_n(\varphi_B)[1\pm\Psi_n(\varphi_D)\Psi_n(\varphi_C)] - \sum_{n=1}^{N}\Psi_n(\varphi_A)\Psi_n(\varphi_C)[1\pm\Psi_n(\varphi_D)\Psi_n(\varphi_B)]\right| \leq \quad (4)$$

$$\frac{1}{N}\sum_{n=1}^{N}[1\pm\Psi_n(\varphi_D)\Psi_n(\varphi_C)] + \frac{1}{N}\sum_{n=1}^{N}[1\pm\Psi_n(\varphi_D)\Psi_n(\varphi_B)] = 2\pm(E_{DC}+E_{DB})$$

It seems that we succeeded to prove Bell-CHSH inequality for estimated pairwise expectations: $|E_{AB} - E_{AC}| + |E_{DB} + E_{DC}| \leq 2$ which may be compared with quantum predictions, but it is not true.

Our proof is only valid, if for each pair of settings (AB), (BC), (DB) and (DC) and in each time slot ΔT we have the same N photon-pairs, which are described by <u>exactly the same set of pairs of functions</u>. It is highly improbable.

We do not know exactly N in each time slot ΔT. N is a random variable. In each setting we may also have different photon-pairs described by different pairs of functions Ψ(φ) thus :

$$E_{AB} = \frac{1}{N_{AB}}\sum_{n=1}^{N_{AB}}\Psi_n(\varphi_A)\Psi_n(\varphi_B) \quad (5)$$

$$E_{AC} = \frac{1}{N_{AC}}\sum_{l=1}^{N_{AC}}\Psi_l(\varphi_A)\Psi_l(\varphi_C) \quad (6)$$

$$E_{DB} = \frac{1}{N_{DB}}\sum_{m=1}^{N_{DB}}\Psi_m(\varphi_D)\Psi_m(\varphi_B) \quad (7)$$

$$E_{DC} = \frac{1}{N_{DC}}\sum_{k=1}^{N_{DC}}\Psi_k(\varphi_D)\Psi_k(\varphi_C) \quad (8)$$

Since
$$E_{AC} = \frac{1}{N_{AC}} \sum_{l=1}^{N_{AC}} \Psi_l(\varphi_A)\Psi_l(\varphi_C) \neq \frac{1}{N_{AB}} \sum_{n=1}^{N_{AB}} \Psi_n(\varphi_A)\Psi_n(\varphi_C) \qquad (9)$$
and $|E_{AB} - E_{AC}| + |E_{DB} + E_{DC}| \leq 2$ may not derived, as it was attempted in (4).
Besides we do not observe and follow photon-pairs, but only register clicks on the distant detectors and it is not an easy and error free task to identify clicks produced by the same pair of photons.

If for each pairs of polarizers we gather a lot of data using many successive time slots, it may happen that the overall set of spin functions, which describe photons in different settings, may have a significant overlap. In this case we may estimate statistical errors for different settings and replace: $|E_{AB} - E_{AC}| + |E_{DB} + E_{DC}| \leq 2$ by:

$$|\bar{E}_{AB} - \bar{E}_{AC} \pm k_1 \Delta \bar{E}_{AB} \pm k_2 \Delta \bar{E}_{AC}| + |\bar{E}_{DB} + \bar{E}_{DC} \pm k_3 \Delta \bar{E}_{DB} \pm k_4 \Delta \bar{E}_{DC}| \leq 2 \qquad (10)$$

where $\bar{E}_{AB} = \bar{M}_{AB}/\bar{N}$ and $|\Delta \bar{E}_{AB}| = |\Delta \bar{M}_{AB}/\bar{N}| + |\Delta \bar{N}(\bar{M}_{AB}/\bar{N}^2)|$ etc. . The experimental statistical error $\Delta \bar{N}$ is estimated during the calibration without polarizers. The statistical errors $\Delta \bar{M}_{AB}$ are estimated using $-N \leq M_{AB} = N_{AB}(equal) - N_{AB}(not\,equal) \leq N$ etc.

Free parameters $k_1,.. k_4$ allow to find $k_{min}$, which is the smallest value of "standard deviations" needed in order that the inequality (10) is not violated. If one neglects photon-pairs identification errors one might estimate $\Delta \bar{M}_{AB}$ using the assumption that $M_{AB} = N(P_{AB}(++) + P_{AB}(--) - P_{AB}(+-) - P_{AB}(-+))$ . Here $M_{AB}$ is not an independent random variable and the probabilities are those predicted by QM.

We see some advantages of our approach. We may control experimental errors all the time, our proof is not using the assumption (3) and it explains why Bell–CHSH may not be rigorously derived, as it was done in (4). <u>Under the assumption of reproducibility of hidden variables</u> one may derive only the inequality (10) and study the influence of experimental errors on the significance of the violation Bell-CHSH inequality by estimates of pairwise expectations obtained using finite experimental samples for different pairs of settings.

Let us now discuss a different experiment. In this experiment we have a stable source sending, with a fixed intensity, photons/electrons towards successive filters. If the passage by a filter for an individual "particle" is predetermined before an experiment is done, then as Wigner demonstrated one may prove Bell-type inequalities for pairwise joint transition probabilities, which may be violated by quantum predictions.

As we mentioned above, Wigner used the integral (3) to prove his inequalities. By using functions $\Psi'(\varphi) = 0$ or $1$, for $\varphi \in [0, \pi]$ describing states of individual particles we may derive Wigner inequalities for estimates of transition probabilities. If $\Psi'(\varphi) = 1$, a particle is transmitted by a filter "$\varphi$", otherwise it is absorbed by it. We study successive groups of approximately $N = I\Delta T$ particles sent by a source during a time interval $\Delta T$. Using Wigner's notation we obtain:

$$P_{ABC^\perp} \approx (++-) = \frac{1}{N}\sum_{n=1}^{N} \Psi'_n(\varphi_A)\Psi'_n(\varphi_B)\Psi'_n(\varphi_{C^\perp}) \tag{11}$$

where $(++-)$ is a fraction of the particles which may pass by 3 successive filters A, B and $C^\perp$.

Similarly a fraction of particles which may pass by two successive filters A and B is:

$$P_{AB} \approx (++\cdot) = \frac{1}{N}\sum_{n=1}^{N} \Psi'_n(\varphi_A)\Psi'_n(\varphi_B) \tag{12}$$

Only if we have <u>the same set of functions in each experiment</u>, we obtain the following obvious equalities:

$$P_{AC^\perp} \approx (+\cdot-) = (++-) + (+--) \tag{13}$$
$$P_{BC^\perp} \approx (\cdot+-) = (++-) + (-+-) \tag{14}$$
$$P_{AB^\perp} \approx (+-\cdot) = (+-+) + (+--) \tag{15}$$

From (13-15) we obtain:

$$P_{AC^\perp} \approx P_{BC^\perp} - (-+-) + P_{AB^\perp} - (+-+) \tag{16}$$

and finally Wigner's inequality:

$$P_{AC^\perp} \leq P_{BC^\perp} + P_{AB^\perp} \tag{17}$$

which is violated by quantum predictions for some angles : $P_{AC^\perp} = \frac{1}{2}\sin^2(\varphi_A - \varphi_C)$ etc.

As we mentioned above we may only derive (13-17), if we have the same functions describing particles produced in all runs of different experiments described above. If we assume reproducibility of hidden variables for large N instead of (17) we may only derive another inequality:

$$P_{AC^\perp} \pm k_1 \Delta P_{AC^\perp} \leq P_{BC^\perp} \pm k_2 \Delta P_{BC^\perp} + P_{AB^\perp} \pm k_3 \Delta P_{AB^\perp} \tag{18}$$

The inequality (18) similarly to the inequality (10) allows finding $k_{min}$, which is the smallest value of "standard deviations" needed in order that the inequality (18) is not violated.

This terminates my handwritten notes, which I gave to John Bell in 1976.

### 3. Bell's letter to Pitowski

The letter, I received in Morocco, contained 4 pages: a few lines of introduction followed by a copy of his letter to Pitowski:

      1982 Sep 7

Dear Dr K

> Did you notice Pitowski's
> paper    Phys.Rev.Lett.48(1982) 1299
>                          ?
>        With best wishes
>              John Bell

CERN, 1982 Sep 6

Dear Dr Pitowski, thank you for your papers. I have much difficulty in understanding your proposal-not least because my competence in mathematics does not go beyond kindergarten. Let me reformulate what I think you might be saying in the language I know. Let A and B ($= \pm 1$) be the possible results on the two sides. Let
$$E(a,b) = \overline{AB}$$
Be the expectation value of the product for a given experimental settings a, b. The locality hypothesis is that there are some variables $\lambda$, and some functions $\overline{A}(a,\lambda), \overline{B}(b,\lambda), \rho(\lambda)$ with
$$|\overline{A}(a,\lambda)|, |\overline{B}(b,\lambda)| \leq 1$$

$$|\rho(\lambda)| \geq 0$$

Such that
$$E(a,b) = \int d\lambda \rho(\lambda) \overline{A}(a,\lambda) \overline{B}(b,\lambda)$$

From this follows the CHSH inequality
$$|E(a,b) - E(a,b')| + |E(a',b) - E(a',b')| \leq 2$$
<u>Provided</u> the operation
$$\int d\lambda \rho(\lambda)$$
is independent of a and b.

In your model $\lambda$ is a rotation matrix α, defined by three parameters which may be chosen in many ways. You seem to have devised functions $\overline{A}(a,\lambda)$ and $\overline{B}(b,\lambda)$ so pathological that the integral is not defined in general, but can be defined by choosing the three parameters, and the order of integration, as a function of *a* and *b*. Then CHSH does not follow.

  <u>But</u>, in simulating a large but finite experiment we are not concerned with integration over, but with <u>sampling</u> in the space $\lambda$. That is to say that what matters is not the existence of the Riemann-, or Lebesque-, or whoever-, integral but of the <u>Monte Carlo</u> integral. Your functions would make Monte Carlo integration- and long experimental running-unreliable in so far as they avoided CHSH.

  This is something that I learned long ago from M. Kupczyński, who considered whether measure theoretic subtleties could be important. I do not know whether he ever published.

  B. d'Espagnat also, in his books, usually assumes the validity of sampling and induction rather than integration.

There is a very good chance that I have misunderstood you completely. I would be very grateful if you could correct me in the kind of language used here.

<div style="text-align: center;">with best wishes<br>John Bell</div>

Copy to  M. Kupczyński
         B.d'Espagnat
         H. Stapp

It took some time before this letter arrived to Morocco. I replied immediately saying, that I had not published my manuscript and that I would read Pitowski's paper. In Rabat, the access to several scientific journals was quite restricted. I had several teaching and academic duties and only in 1984 during my two week stay in ICTP in Trieste I finalized a short article about completeness of QM, Pitowski's model and BI [27].

## 4. My reply to John Bell and subsequent publications.

In [27], I explained that in the theory with supplementary parameters each pure quantum ensemble is mixed with respect to these parameters and this can be tested using statistical non parametric compatibility tests similar to the purity tests [16-18]. I also reformulated Pitowski's model showing, that definite values of spin functions in all directions, did not necessarily predetermine values of measured spin projections. Finally, I pointed out, that a violation of Bell's inequalities, did not imply a violation of Einsteinian locality because the inequalities were derived by assuming that measured probabilities in incompatible experiments might be determined by conditionalization from a unique probability space, what not always was justified.

In conclusion I wrote: "the theoretical and experimental analysis of EPR paradox and of Bell's inequalities imposed serious restrictions on the models with supplementary parameters and showed that they have to respect in some way Bohr's idea of complementarity… we hope that results of the purity tests we proposed above will give a new comprehensive answer to the EPR-question concerning the completeness of quantum mechanics". In Poland we have a good proverb: "hope is a mother of fools".

The paper was submitted to Physical Review Letters (PRL) and peer reviewed. A reviewer suggested splitting it into 3 parts and resubmitting them separately. Being politely saying naïve, I refused, asked for an adjudicator and finally my paper was rejected.

In June 1985, I participated in the Symposium on the Foundations of Modern Physics in Joensuu: 50 years of the Einstein-Podolsky-Rosen-Gedankenexperiment. It was a very important and well organized conference. Among participants were: Aerts, Barut, Beltrametti, Bub, Bush, Enz, Van Frassen, Grangier, Horne, Ingarden, Jammer, Kochen, Kraus, Lahti, Mittelstaedt, De Muynck, Ne'man, Peierls, Piron, Prugovecki, Pykacz, Randal, Rayski, Rohrlich, Rosen, Stapp, von Weizsäcker and Zeilinger.

After reading Symposium Proceedings [28], I discovered that several authors: Aerts [29,30], Angelidis and Poppper [31] disagreed with the common interpretation of the violation of Bell's inequalities and as Barut [32] concluded: " *one has to be cautious in drawing-far-reaching philosophical conclusions from Bell's inequalities"* .

I also realized that the referee in PRL was right. I divided [27] into 3 parts, I added a lot of new material and I submitted three manuscripts to Physics Letters A, where they were published [33-35], without any problem. In [34], I explained Bertrand's Paradox and an intimate relation of probabilistic models with experimental protocols. In particular, I pointed out that experimental protocols in spin polarisation correlation experiments (SPCE) were inconsistent with probabilistic models used to derive Bell's inequalities. In local realistic hidden variable model experiments performed in incompatible measurement settings were described using jointly distributed random variables on a unique probability space and conditionalization. Therefore, the violation of inequalities had nothing to say about Einsteinian locality. In [33-35] I cited only articles which I knew.

Several other authors came earlier to similar conclusions [36-41], but I discovered it many years later. I don't know, whether Pitowski replied to Bell's letter, probably his article about Boole's inequalities [41] was meant to be his reply.

I sent a copy of my papers [33-35] to Bell, but probably I did not convince him. He never answered or perhaps his letter arrived to Morocco, when I was gone. Unfortunately Bell passed away in 1990.

For me the topic of Bell inequalities and their violation was well understood and closed. During the Symposium on Mathematical Physics held in Toruń in 2001, Andrzej Kossakowski told me about a preprint of Accardi and Regoli [42], in which they presented results of a computer experiment violating Bell's inequalities and gave several arguments that there was no contradiction between quantum theory and locality.

After reading this preprint, I realized, with surprise, that there was still no consensus about the violation of Bell inequalities. Instead of rejoicing that there was not, necessarily, contradiction between QT and Einsteinian locality many members of physical community seemed to accept that two perfect random dices in distant locations might give perfectly correlated outcomes. They did not understand that the entanglement and the violation of Bell inequalities did not justify such a 'physical picture'.

Evoking quantum magic is counterproductive and misleading this is why I got back to work on the subject. I started to attend conferences and to write articles [6, 43-46]. Since 2007, I have attended regularly Conferences on the Foundations of Quantum Mechanics organized every year in Växjö by Andrei Khrennikov. During these conferences I met several colleagues, who arrived, very often independently, to similar conclusions about BI-CHSH and their violation.

We tried to convince quantum information community (QIC) and experimentalists, that metaphysical implications of the violation of inequalities were quite limited [47-101]. 'Photon pairs' in Bell Tests could not be described as pairs of socks or pairs of dice and proposed hidden variable probabilistic models were inconsistent with experimental protocols used in various Bell Tests. Additional arguments, against quantum non-locality, were given in [102-106] and the most recently in [107].

In spite of this, quantum mysteries are still the topic of choice not only on social media but also in scientific articles. It seems that *magic sells better* [108].

## 4. Conclusions

In his letter to Pitowski, Bell derived CHSH inequality using $\overline{A}(a,\lambda), \overline{B}(b,\lambda)$ obtained by averaging over some <u>unspecified</u> variables. For example, it could be variables describing measuring instruments as in [83,109]. Such averaging cannot be implemented in SPCE and even if it could, it would destroy all correlations created by a source. This is why using a language of mathematical statistics the equation:

$$E(a,b) = \int d\lambda \rho(\lambda) \overline{A}(a,\lambda) \overline{B}(b,\lambda) \qquad (19)$$

called by Bell *locality hypothesis* describes a family of pairs of independent random experiments parametrized by $\lambda$. A factorisation in (19) means only a statistical independence of corresponding random variables for each value of $\lambda$. To estimate $E(a, b)$, for each $\lambda$ two independent random experiments are repeated and $\overline{A}(a,\lambda), \overline{B}(b,\lambda)$ are calculated; next these calculated values are averaged over all $\lambda$. This experimental protocol resembles a protocol implied by a stochastic hidden variable model (SHVM), but it has nothing to do with experimental protocols used in Bell Tests. A detailed discussion of various experimental protocols and their intimate relation to probabilistic models may be found in [75].

Not only the integrals (3) and (19) are incompatible with the experimental protocols but in general they do not exist, if $A(a,\lambda), B(b,\lambda)$ are spin functions similar to those discussed by us [34,35] and by Pitowski [25,26]. Thus, the averaging over instrument variables under the integral sign as in [110] is meaningless.

Bell insisted, that $\rho(\lambda)$ could not depend on experimental settings, because he incorrectly believed that $\rho(\lambda) \neq \rho(\lambda | a,b)$ would imply super-determinism. Only many years later it was explained, that such belief was unfounded and based on a questionable use of Bayes Theorem and incorrect causal interpretation of conditional probabilities [79, 85-88]. Therefore, setting dependence of hidden variables should be called *contextuality* instead of *conspiracy* or *super-determinism.*

In fact, if one wants to use hidden variables, $\rho(\lambda)$ has to depend on settings, because in Bell Tests and in several experiments in cognitive science the 'marginal laws' called *no-signaling* are also violated [69,80,86, 110-119]. Therefore, these experiments can neither be explained by non-contextual hidden variable model LRHVM nor by quantum probabilistic model describing an ideal EPRB experiment.

Bell was a reasonable man and working in CERN, he never believed that *experimenters' freedom of choice* might be compromised. He would never agree, that *an electron or proton can be here and a meter away at the time* or that the violation of his inequalities proves that: *two perfectly random distant events produce always perfectly correlated outcomes*. In our opinion, he would probably now choose *contextuality* instead of *nonlocality* [88].

Meaning of BI-CHSH and their violation is now well understood. These inequalities imposing the bounds on particular cyclic combinations of pairwise expectations $E(a, b)$, of random variables can be only rigorously proven for random experiments outputting 4 outcomes ±1 in each trial, which can be described by a probabilistic model in which there exists a well-defined joint probability distribution of the corresponding 4 random variables. In such experiments BI-CHSH are strictly obeyed by the estimates of *E (a, b)* obtained using finite samples of any size. In Bell Tests, for each pair of settings (a, b) only pairs of outcomes ±1 are outputted and one can only investigate the plausibility of various probabilistic couplings [88]. The estimates of pairwise expectations obtained using a finite samples may significantly violate BI-CHSH. In particular one can reject with high confidence LRHVM and SHVM models/couplings but this does not allow for far reaching metaphysical conclusions about quantum non-locality, retro-causality or super-determinism[53,65,69,79,82,88,105-107].

As Hans de Raedt et al. clearly explained in [107]: " all EPRB experiments which have been performed and may be performed in the future and which only focus on demonstrating a violation BI-CHSH merely provide evidence that not all contributions to the correlations can be reshuffled to form quadruples… These violations do not provide a clue about the nature of the physical processes that produce the data….in contrast to Bell's original derivation, the derivation of Bell-type inequalities in the probabilistic setting does not rely on assumptions about "locality", "macroscopic realism", "non-invasive measurements" and the like. Violations of Bell-type inequalities derived within the framework of a probabilistic model are a signature of the non-existence of a joint distribution rather than of some signature of "quantum physics". Most importantly, describing two-valued data of EPRB experiments performed under four different conditions in terms of a joint distribution (if it exists) accomplishes exactly the opposite of the description in separated parts provided by quantum theory".

At the end of this review, I include the conclusions from [6], which in my opinion did not lose their actuality. I added only some additional references.

"The statistical interpretation of QM [45, 46, 54, 61-62, 79, 81, 120-123] is consistent. Experimental tests of Bell's theorem can neither confirm a completeness of QM, nor prove that the only models with supplementary parameters able to give "a microscopic description" of SPCE have to violate Einstenian causality.

A question whether a statistical description provided by the quantum theory gives a complete description of experimental data is fully justified. This question cannot be answered by proving a mathematical theorem or by constructing ad hoc models with supplementary parameters reproducing some quantum predictions. It can be only answered by a detailed analysis of time- series of experimental results, which can be done with the help of purity tests, which were proposed many years ago and never done. If deviations from randomness were detected and some new regularity found, the standard statistical description given by the quantum theory should be completed by a description using probably ideas of stochastic processes.

In some sense, this change of description has already been made in stochastic approaches used to explain various phenomena involving trapped atoms, ions and molecules. In these approaches, a wave function obeys a Schrödinger equation with an effective Hamiltonian separated by quantum jumps occurring at random times. Purity tests could be used to check these new stochastic models, which assume without checking *ergodicity* of observed time-series. A question about completeness of QM formulated in this way is independent of existence or non- existence of a detailed 'microscopic description' of phenomena presenting this particular stochastic behavior.

From Bertrand's paradox we learned, that we should not talk about the probabilities without referring to random experiments used to estimate them. Therefore, the quantum theory providing probabilistic predictions should not lose its contact with experiments, it wants to describe.

If one forgets that QM does not give any 'microscopic images', and provides only mathematical algorithms able to describe statistical regularities observed in the data, one creates incorrect subquantum models which lead to paradoxes and speculations, which seem to be a pure science fiction [81].

Quantum observables are contextual, what means that their values are not attributes of individual members of a quantum ensemble, but they give only information about the possible interactions of a whole ensemble of identically prepared physical systems with measuring devices. If ensembles are pure, one can speak about probabilistic information pertinent to the interaction of each individual system with a measuring device. To be able to do it, one must check the purity of prepared quantum ensembles. There is no strict anti-correlations between two time-series of outcomes obtained in SPCE, thus these two time series may not be used in quantum cryptography to create a secret key, shared by Alice and Bob without an appropriate error correction protocol.

The purity tests are important and relatively simple [16-18, 33, 70-71], data are available. We hope that this paper will convince some experimentalists to use them."

Perhaps this time the polish proverb: "hope is a mother of fools" will turn to be wrong.

**Funding:** This research received no external findings.
**Institutional Review Board Statement:** Not applicable.
**Informed Consent Statement:** Not applicable.
**Data Availability Statement:** Not applicable.